\documentclass[12pt,a4paper]{article}

\usepackage{a4}
\usepackage{amsmath}
\usepackage[dvips]{graphics}
\usepackage{pstricks}
\usepackage[square, numbers]{natbib}
\usepackage{latexsym}

\begin{document}

\title{\bfseries An early sign of satisfiability}
\author{Eliezer L. Lozinskii \\
School of Computer Science and Engineering \\
The Hebrew University, Jerusalem 91904, Israel \\
{\it email: lozinski@cs.huji.ac.il}}
\date{}

\maketitle

\newtheorem{definition}{Definition}[section]
\newtheorem{theorem}{Theorem}[section]
\newtheorem{example}{Example}[section]
\newtheorem{lemma}{Lemma}[section]
\newtheorem{corollary}{Corollary}[section]
\newtheorem{conjecture}{Conjecture}[section]
\newtheorem{proposition}{Proposition}[section]
\newtheorem{algorithm}{Algorithm}[section]
\newtheorem{observation}{Observation}[section]

\begin{abstract}

This note considers checking satisfiability of sets of propositional 
clauses (SAT instances).
It shows that \emph{unipolar sets} of clauses (containing no 
positive or no negative clauses) provide an \emph{early sign} of 
satisfiability of SAT instances
{\bfseries before} all the clauses become satisfied in the course 
of solving SAT problems. At this sign the processing can be terminated
by \emph{unipolar set termination, UST}
thus before it is usually done by SAT solvers (Table 1). An analysis of
benchmark SAT instances used at SAT Competitions shows that UST can
speed up solving SAT instances stemming from many real-world problems.
The efficiency of UST increases with the \emph{skewnwss} of the SAT set being 
checked, that is the difference between probabilities of negated and unnegated 
literals in the set. 
Many real-world problems, by virtue of their semantics, are skewed (Table 2). 
The efficiency of UST can be increased by revealing the \emph{hidden skewness} 
of SAT sets (Table 3).

\end{abstract}


\section{Introduction}

\emph{SAT} is the  problem of checking satisfiability of a propositional 
formula $F$ presented in the conjunctive normal 
form $CNF$ as a conjunction of a set
of \emph{clauses}; each clause is a disjunction of \emph{literals},
each literal is an unnegated or negated propositional \emph{variable}.
Let $S$ be the set of clauses of $F$, and $V$ --- the set of variables 
appearing in $S$. In the process of checking satisfiability of $F$ each
variable $v \in V$ can be assigned a truth 
value \emph{true} or \emph{false}.
Let $A$ be a set of assignments to the variables of $V$. A clause $C \in S$
is \emph{satisfied} if there is a literal in $C$ assigned \emph{true}
in $A$.
$F$ is \emph{satisfiable} if there is a set of assignments that
satisfies all the clauses of $S$, otherwise $F$ is \emph{unsatisfiable}.

Solvers for SAT are based on the famous 
Davis-Putnam-Logemann-Loveland procedure \emph{DPLL} \citep{dp60, dav62}.
In search for a set of assignments $A$ satisfying all the clauses of $S$
DPLL performs a sequence of \emph{steps}. At each step it assigns a truth 
value to a variable $v \in V$ not yet assigned, and \emph{applies} this
assignment $\alpha$ to $S$ in the following way: for every clause $C$ of $S$
(not yet deleted from $S$), if $C$ contains a literal assigned \emph{true} by $\alpha$
then $C$ is deleted from $S$, if $C$ contains a literal $L$ assigned 
\emph{false} by $\alpha$ then $L$ is deleted from $C$; if after deleting 
$L$ from $C$ the latter becomes empty then DPLL encounters a 
\emph{conflict}, so it \emph{backtracks} 
cancelling the changes produced by the application of $\alpha$
and then performing another assignment instead of $\alpha$; otherwise
if the application of $\alpha$ causes no conflict then the assignment is 
\emph{successful} and $\alpha$ is appended to $A$;
DPLL proceeds to the next step; if after deleting $C$ from $S$ 
the latter becomes empty (meaning that all clauses of $S$ are satisfied 
by $A$) then DPLL terminates deciding that $S$ 
is \emph{satisfiable}; however, if the search is exhausted, but a 
satisfying set of assignments is not found then $S$ is 
\emph{unsatisfiable}.

In the course of the fifty-odd years since DPLL was pioneered numerous 
innovations have been introduced and engineered into the procedure, so 
that the modern SAT solvers are sophisticated and efficient programs 
capable of checking satisfiability of very large sets containing 
millions of clauses (a profound description and analysis of SAT solvers 
is presented in \citep{hand, gom08}.
Modern solvers use efficient branching strategies 
for determining next assignment; they perform long backtracking jumps 
pruning large portions of the search space; when conflicts are 
encountered, new \emph{learned clauses} are constructed and added 
to the set to prevent repeated assignments leading to the same conflict; 
to reduce the amount of book-keeping many solvers using a \emph{lazy} 
data structure do not update the state of all literals of every clause, 
but keep track of certain \emph{watched literals}; these solvers 
determine satisfiability of a set of clauses when all its variables 
have been assigned successfully \citep{marq09}
(at this stage all clauses of the set are satisfied). 

Solvers fully implementing DPLL are \emph{complete} such that for any 
set of clauses they decide whether it is satisfiable or not \citep{dar09}.
There are solvers that unlike complete ones do not carry 
out an exhaustive search of a satisfying assignment, but perform a 
stochastic \emph{local search} with no guarantee of finding an existing 
one \citep{kau09}.
Although the solvers based on local search often outperform 
complete solvers, they are \emph{incomplete} since for unsatisfiable 
sets and even for some satisfiable ones these solvers terminate undecided.

\section{Unipolar sets}

Consider a set of clauses $S$ that is being checked for satisfiability 
by a SAT solver. In the course of this process the contents and size of 
$S$ change. At any step when an assignment $\alpha$ is applied to $S$, 
clauses satisfied by $\alpha$ are deleted from $S$ or marked as 
\emph{inactive} (depending on the working solver), so the set of 
\emph{active} clauses $\tilde{S}$ subject to the further processing 
shrinks. On the other hand, when a conflict (an empty clause) is 
encountered and so the search backtracks, some inactive or previously 
deleted clauses may be returned into $\tilde{S}$ extending it. 

Such a ``pulsating'' behaviour is characteristic of any set of clauses 
processed by any of a wide variety of solvers. And important: no matter 
what is the method of making assignments, backtracking or book-keeping 
implemented by a solver --- when the solver terminates reporting 
satisfiability of a set $S$, then there is an assignment found by the 
solver that satisfies all the clauses of $S$ (although solvers using a 
lazy data structure \citep{marq09}
may be unaware of this fact). Let us call this 
event the \emph{all-satisfied termination, AST}. 

\begin{definition}
If a clause $C$ contains unnegated literals only then call $C$ a
\emph{positive clause}; if $C$ contains negated literals only then
$C$ is a \emph{negative clause}; otherwise $C$ is a \emph{mixed clause}.
If a set $S$ contains both a positive and a negative clauses then 
call $S$ a \emph{bipolar set}; otherwise $S$ is an \emph{unipolar set}.
$\Box$
\end{definition}

\begin{observation}
Any unipolar set $S$ of clauses is satisfiable. Indeed, if $S$ contains
no positive (negative) clauses then every clause of $S$ contains a negated
(unnegated) literal. So the assignment of \emph{false} (\emph{true}) 
to all variables of $S$ satisfies all clauses of $S$. $\Box$
\end{observation}

By Observation 2.1, the unipolarity of a set of clauses is a sign of 
its satisfiability, so if in the course of checking satisfiability 
of a set $S$ the 
set of active clauses $\tilde{S}$ becomes unipolar, then at this moment 
the search can be terminated deciding that $S$ is satisfiable. 
Let us call this event the
\emph{unipolar set termination, UST}. When UST occurs, $\tilde{S}$ 
is not empty containing clauses that are not yet satisfied by the 
previous assignments, but are known to be satisfiable by assigning 
\emph{true} or \emph{false} to all yet unassigned variables of the set. 
As the steps of variable assignments are performed in
a sequence, let $\alpha_i$ be the truth value assignment to a variable
at step $i$. Let $\tilde{S}_{bi}$ and $\tilde{S}_{ai}$ denote, respectively, 
the states of $\tilde{S}$ immediately before and after application of 
$\alpha_i$.

\begin{proposition}
For all satisfiable sets $S$ of clauses and all sequences of 
variable assignments, 
$\tilde{S}$ becomes unipolar \emph{before} all clauses of $S$ are 
satisfied, so UST is \emph{always achieved in less steps than AST}.
\end{proposition}

{\bfseries Proof}.
Let $\alpha_t$ be the assignment of \emph{true} to a literal $L$ such
that all clauses of $\tilde{S}_{bt}$ become satisfied. So $S$ is 
satisfiable, and 
AST takes place after performing $t$ steps of variable assignment. 
Consider the content of $\tilde{S}_{bt}$
just before the application of $\alpha_t$: $\tilde{S}_{bt}$ is not empty 
(otherwise
AST would have occurred before step $t$); all clauses of $\tilde{S}_{bt}$
are satisfied by $\alpha_t$ so every clause of $\tilde{S}_{bt}$ contains $L$; 
but a positive and a negative 
clauses cannot contain the same literal, so $\tilde{S}_{bt}$ is unipolar:
either no positive or no negative clauses. Should $\tilde{S}_{bt}$ 
be checked for 
unipolarity, satisfiability of $S$ would be detected, and UST performed
before step $t$,
so step $t$ would not be needed. $\Box$

Unipolarity is a common feature of satisfiable sets of clauses 
in the following 
sense. Let $\theta$ denote a subset of variables appearing in a set of 
clauses $S$, and $S\theta$ stand for a set resulting from inverting 
all literals in $S$ involving the variables in $\theta$. Call $\theta$ an
\emph{inverter}.

\begin{proposition}
A set of clauses $S$ is satisfiable iff there exists an inverter $\theta$ 
such that $S\theta$ is unipolar. 
\end{proposition}

{\bfseries Proof}. 
\emph{If:} For all sets $S$ and all inverters $\theta$, $S$ 
is satisfiable iff $S\theta$ is so. Indeed, if $M$ is a model of $S$
then $M\theta$ is a model of $S\theta$, and vice versa. So if $S\theta$
is unipolar then, by Observation 2.1, $S$ is satisfiable.

\emph{Only if:} If $S$ is satisfiable, and $M$ is one of its models,
define an inverter $\theta$ as the set of all unnegated variables in $M$.
$M\theta$ contains negated literals only and is a model of $S\theta$, hence,
$S\theta$ contains no positive clause and so is unipolar. $\Box$

\section{The gain of UST over AST}

To evaluate the efficiency of the UST relative to the AST we have run
experiments with a program that implements DPLL: given a set of caluses $S$, 
the program at each step assigns \emph{true} to the most frequent literal
of a most frequent variable among the active clauses;
after applying every 
variable assignment the program updates the size of the sets of active 
positive and 
negative clauses, and if one of them is empty then at this step the 
processing could be terminate by UST, but the search goes on until all 
clauses of the set are satisfied, so AST is performed. Let $N_U$ and 
$N_A$ denote the number of variable assignments made by the program till 
UST and AST, respectively, are reached, and $G = N_A / N_U$ stand for the 
\emph{gain} of UST over AST. By Proposition 2.1, $N_U < N_A$, so $G > 1$.
Let $R$ denote the \emph{remainder}, that is the percentage of the clauses 
of $S$ remained active but not yet satisfied at UST; $R > 0$. The larger 
the values of $G$ and $R$, the more efficient UST is for $S$.
The sets $S$ in the experiments were generated with the following 
parameters (Table 1): number of variables $n = 100$;
the clauses-to-variables ratio $r = m / n$ varied from $r = 2$
to the threshold value (shown in boldface in Table 1)
at which satisfiability of $S$ undergoes phase transition 
\citep{cra96};
3 literals in every clause; all variables appear
in $S$ with the same probability, however negated and unnegated
literals have different probabilities; without loss of generality,
the probability of an unnegated literal $p \le 0.5$; 1000 instances
were checked for each pair of values $(p, r)$.

Although sets for testing SAT solvers often
are generated with $p = 0.5$, sets with $p < 0.5$ are
theoretically interesting and practically important (next section).
Sets with different probability of negated and unnegated literals
(\emph{skewed} sets) were studied by Sinopalnikov \citep{sin04}, and 
shown to
undergo satisfiability phase transition at a threshold value of $r$
that grows with decreasing value of $p$. The efficiency of UST depends 
on the value of $p$. Indeed, the smaller the value of $p$
in $S$, the smaller the probability that a positive clause appears in $S$
and so larger the probability that $S$ becomes unipolar much earlier
than all its clauses become satisfied in the process of checking its 
satisfiability.

\begin{table}
\caption{The gain $G$ and remainder $R$ for $p = 0.5  - 0.05$, $n = 100$, 1000 tests for each pair $(p, r)$}
\begin{center}
\begin{tabular}{*{12}{|c}|}
\hline
$p$ & $r$ & 2.00 & 2.25 & 2.50 & 2.75 & 3.00 & 3.25 & 3.50 & 3.75 & 4.00 & {\bfseries 4.26} \\
\cline{2 - 12}
0.5 & $G$ & 1.08 & 1.07 & 1.07 & 1.06 & 1.05 & 1.02 & 1.01 & 1.00 & 1.00 & 1.00 \\
\cline{2 - 12}
& $R\%$ & 3 & 3 & 3 & 2 & 2 & 2 & 2 & 1 & 1 & 1 \\
\hline\hline
$p$ & $r$ & 2.00 & 2.30 & 2.60 & 2.90 & 3.20 & 3.50 & 3.80 & 4.10 & 4.40 & {\bfseries 4.70} \\
\cline{2 - 12}
0.4 & $G$ & 1.11 & 1.09 & 1.08 & 1.07 & 1.06 & 1.03 & 1.01 & 1.00 & 1.00 & 1.00 \\
\cline{2 - 12}
& $R\%$ & 4 & 3 & 3 & 3 & 2 & 2 & 2 & 2 & 1 & 1 \\
\hline\hline
$p$ & $r$ & 2.00 & 2.50 & 3.00 & 3.50 & 4.00 & 4.50 & 5.00 & 5.50 & 6.00 & {\bfseries 6.40} \\
\cline{2 - 12}
0.3 & $G$ & 1.22 & 1.18 & 1.15 & 1.12 & 1.10 & 1.05 & 1.02 & 1.00 & 1.00 & 1.00 \\
\cline{2 - 12}
& $R\%$ & 8 & 6 & 5 & 4 & 3 & 3 & 2 & 2 & 2 & 1 \\
\hline\hline
$p$ & $r$ & 2.00 & 3.00 & 4.00 & 5.00 & 6.00 & 7.00 & 8.00 & 9.00 & 10.00 & {\bfseries 11.5} \\
\cline{2 - 12}
0.2 & $G$ & 1.85 & 1.54 & 1.38 & 1.30 & 1.23 & 1.15 & 1.07 & 1.02 & 1.01 & 1.01 \\
\cline{2 - 12}
& $R\%$ & 33 & 21 & 13 & 9 & 7 & 6 & 4 & 3 & 3 & 2 \\
\hline\hline
$p$ & $r$ & 2.00 & 3.00 & 5.00 & 10.0 & 15.0 & 20.0 & 25.0 & 30.0 & 35.0 & {\bfseries 41.0} \\
\cline{2 - 12}
0.1 & $G$ & 8.60 & 6.36 & 4.33 & 2.44 & 1.89 & 1.58 & 1.26 & 1.07 & 1.06 & 1.06 \\
\cline{2 - 12}
& $R\%$ & 76 & 72 & 69 & 45 & 31 & 25 & 16 & 12 & 9 & 9 \\
\hline\hline
$p$ & $r$ & 2.00 & 10.0 & 30.0 & 50.0 & 70.0 & 90.0 & 110 & 130 & 150 & {\bfseries 165} \\
\cline{2 - 12}
0.05 & $G$ & 48.8 & 14.8 & 5.47 & 3.28 & 2.24 & 1.46 & 1.32 & 1.31 & 1.31 & 1.30 \\
\cline{2 - 12}
& $R\%$ & 98 & 91 & 74 & 58 & 48 & 41 & 34 & 33 & 33 & 32 \\
\hline
\end{tabular}
\end{center}
\end{table}

Results of the experiments are summarised in Table 1.
While for $p \ge 0.3$ UST
gains just several percents of assignments over AST, for $p < 0.3$
UST requires many times less assignments than AST. For instance,
for $n = 100, p = 0.1, r = 5.0$, UST is 4.33 times faster than AST: 
UST detects satisfiability after 15
assignments; at this step there remain 345 active clauses 
not yet satisfied (out of 500 
initially), and then AST requires 50 more assignments (averaged over
1000 instances). It is remarkable that satisfiability of a set can 
be decided when a significant part of its clauses (69\% in this example) 
have not been satisfied. 

\section{``Skewness'' of real-world SAT problems}

SAT is important not only in the theory of computation being the core 
NP-complete problem \citep{coo71},
but it has numerous practical applications, 
since many real-world 
problems can be encoded as SAT instances. Among these SAT encodings are 
such 
important problems and techniques as planning, digital circuits design, 
software verification, network design and analysis, diagnosing, data 
security, cryptanalysis, graph colouring, proof checking, automated 
reasoning \citep{bry99, cla01, een06, kau96, ste96, vel03, viz15}.

SAT encodings stemming from practical problems are likely to be ``skewed'', 
since unnegated and negated literals represent different aspects of the 
encoded 
problems. For example, unnegated literals usually represent certain 
features of 
the corresponding objects, while negated literals participate in 
encoding of 
restrictions and conditions imposed upon these features. The restrictions 
involve usually combinations of features and are often more numerous than 
individual features and objects. So by virtue of their semantics many
practical problems produce skewed SAT instances. 

\begin{table}
\caption{Skewness parameters of practical SAT instances}
\begin{center}
\begin{tabular}{|c|l|c|c|}
\hline
{\bfseries Source} & {\bfseries Family} & {\bfseries \# inst} & {\bfseries p} \\
\hline
& Planning blocksworld & 7 & 0.14 - 0.26 \\
\cline{2 - 4}
SATLIB & Planning logistics & 4 & 0.28 - 0.42 \\
\cline{2 - 4}
\citep{LIB} & Graph colouring & 400 & 0.08 - 0.13 \\
\cline{2 - 4}
& Quasigroup encoded & 22 & 0.003 - 0.278 \\
\hline
SAT-2005 & Maris & 67 & 0.03 - 0.33 \\
\cline{2 - 4}
\citep{comp} & Grieu & 12 & 0.24 \\
\hline
SAT-2014 & Oldpool & 8 & 0.15 - 0.40 \\
\cline{2 - 4}
\citep{sc14} & Wallner & 20 & 0.07 - 0.15 \\
\hline
\end{tabular}
\end{center}
\end{table}

Table 2 shows skewness parameters of several benchmark 
SAT instances of the category \emph{industrial} and \emph{applications} 
from SATLIB \citep{LIB} and SAT Competitions 
\citep{sc14, comp}
(540 instances). 
For almost all these instances, their values 
of {\bfseries p} fall within the range of a significant gain
of UST (Table 1). These findings suggest that performing UST 
\emph{instead of}
AST would speed up practical SAT solvers. Notably, all 13 instances of the
subset \emph{/wallner-argumentation/complete} from SAT 2014 competition 
(part of the last line of Table 2) are unipolar initially 
(no positive clauses), so by UST the processing would be terminated
at the very first step ``in no time''. This fact went by unnoticed such
that these instances appeared again as benchmarks at the SAT Race 2015
Competition (www.baldur.iti.kit.edu/sat-race-2015/index.php?cat=downloads).

Although UST requires less assignments than AST, it involves more 
book-keeping for updating the numbers of positive and negative active clauses.
The efficiency of UST depends strongly on the skewness of the set. 
In order to decide whether to perform UST for a given set
(or spare this additional book-keeping for a set with the value of {\bfseries p} 
close to $0.5$) the value of {\bfseries p} 
can be counted in linear run-time at a preprocessing. For significantly 
skewed SAT sets, UST is much faster than AST, so for this kind of SAT 
instances corresponding to many practical problems,
all solvers, complete and incomplete, can benefit from 
performing UST.

\section{Hidden skewness}

The initial skewness of a SAT instance can be enhanced without affecting
its satisfiability, so increasing the efficiency of UST.
Let $pos(v, S)$, $neg(v, S)$ denote, respectively, the number of 
unnegated, negated occurrences of a variable $v$ in a set of clauses $S$,
and $poslit(S)$, $neglit(S)$ stend for the total number of unnegated,
negated literals in $S$. Then the skewness $p(S)$ of $S$ is
$$
p(S) = min(poslit(S), neglit(S))/(poslit(S) + neglit(S)).
$$

\begin{definition}
Given a set $S$, define an inverter $\rho_S = \{v \: | \: pos(v, S) > neg(v, S)\}$.
We say that $\rho_S$ \emph{reveals} the \emph{hidden skewness} of $S$, $hp(S)$,
such that $hp(S) = p(S\rho_S)$ $\Box$
\end{definition}

\begin{proposition}
For all sets $S$, $hp(S) \le p(S)$.
\end{proposition}

{\bfseries Proof}.
By inverting in $S$ all literals involving the variables of $\rho_S$ we get:
\newline
for all $v \in \rho_S$,
\newline
$pos(v, S\rho_S) < neg(v, S\rho_S),\:
pos(v, S\rho_S) < pos(v, S),\: pos(v, S\rho_S) = neg(v, S),$ 
\newline
while for all $v \not\in \rho_S$, 
\newline
$pos(v, S\rho_S) \le neg(v, S\rho_S),\:
pos(v, S\rho_S) = pos(v, S),\: pos(v, S\rho_S) \le neg(v, S).$ 
\newline
Hence, 
\newline
$poslit(S\rho_S) \le neglit(S\rho_S),\:
poslit(S\rho_S) \le poslit(S),\:
poslit(S\rho_S) \le neglit(S),$ 
\newline
and so $p(S\rho_S) \le p(S)$. $\Box$

Table 3 shows hidden skewness of a sample of benchmark instances 
from the SAT Race 2015 and SAT 2016 Competitions 
(www.baldur.iti.kit.edu/(sat-race-2015 and sat-competition-2016)/index.php?cat=downloads). 
For many SAT sets $hp(S)$ is significantly less than $p(S)$, so revealing 
the hidden skewness of a given set increases the efficiency of UST. Inversion
of literals involving the variables of $\delta$ can be performed at a
preprocessing in a linear run-time.

\begin{table}
\caption{Hidden vs initial skewness of benchmark SAT instances}
\begin{center}
\begin{tabular}{|c|c|c|c|c|c|}
\hline
{\bfseries Source} & {\bfseries Instance (File)} & {\bfseries \#var} & {\bfseries \#cla} & {\bfseries Initial} & {\bfseries Hidden} \\
& {\bfseries S} & {\bfseries n} & {\bfseries m} & {\bfseries p(S)} & {\bfseries hp(S)} \\
\hline
& jgiraldezlevi.2200.9086. & & & & \\
& 08.40.108.cnf & 2200 & 9086 & 0.499 & 0.388 \\
\cline{2 - 6}
SAT-Race-2015 & manthey\_single-ordered- & & & & \\
\citep{sr15} & initialized-w18-b8.cnf & 2160 & 15054 & 0.407 & 0.364 \\
\cline{2 - 6}
& partial-5-17-s.cnf & 252328 & 1189896 & 0.407 & 0.359 \\
\cline{2 - 6}
& aaai10-planning-ipc5- & & & & \\
& TPP-21-step11.cnf & 99736 & 783991 & 0.352 & 0.260 \\
\hline
& C168\_FW\_UT\_518.cnf & 1909 & 7511 & 0.446 & 0.234 \\
\cline{2 - 6}
SAT-2016 & gripper14u.cnf & 4584 & 43390 & 0.249 & 0.172 \\
\cline{2 - 6}
\citep{sc16} & korf-18.cnf & 7794 & 186934 & 0.169 & 0.027 \\
\cline{2 - 6}
& E00N23.cnf & 15364 & 2210893 & 0.150 & 0.004 \\
\hline
\end{tabular}
\end{center}
\end{table}

\begin{example}
Given a set $S = \{(v_!, v_2, v_3), (\neg v_1, \neg v_2, \neg v_3), (\neg v_1, v_2, \neg v_3)\}$, we get: $p(S) = 0.44, \rho_S = \{v_2\}$,
$S\rho_S = \{(v_1, \neg v_2, v_3), (\neg v_1, v _2, \neg v_3), (\neg v_1, \neg v_2, \neg v_3)\}$ and the hidden skewness of $S$, $hp(S) = p(S\rho_S) = 0.33$. By the way, $S\rho_S$ is already unipolar. $\Box$
\end{example}

\section{Conclusion}

The \emph{unipolar set termination}, \emph{UST}, is presented
that detects satisfiability in the process of solving SAT 
\emph{always before} the all-satisfied termination, AST, 
performed usually by SAT solvers, takes place.

We measure the efficiency of UST by its \emph{gain}, $G$, that is the
ratio of the number of variable assignments required by AST to that of UST.
Table 1 shows values of $G$ produced by experiments with SAT instances
generated with varied parameters.
UST is most efficient for \emph{skewed} sets of clauses with different
probability of unnegated and negated literals. It turns out that SAT
instances encoding real-world problems are very likely to be significantly
skewed. This is true of many instances stemming from practical problems used
as benchmarks for SAT competitions (shown in Tables 2, 3). 
Implementing UST requires updating the number of positive and negative 
active clauses at every variable assignment, but as suggested by this 
short study, for many important practical SAT problems the gain of UST 
over AST can by far compensate for this additional book-keeping, thus 
speeding up the process of solving SAT. Revealing the hidden skewness 
of a SAT sets (Table 3) can increase the efficiency of UST.

\bibliographystyle{plain}

\end{document}